\documentclass[final,5p,times,sort&compress, fleqn]{elsarticle}

\usepackage{amssymb}
\usepackage{amsthm}
\usepackage{natbib}

\usepackage{multirow}
\usepackage{amsmath}
\usepackage{multicol}

\journal{Physica E}

\begin{document}

\begin{frontmatter}

\title{Magnetization plateaus and bipartite entanglement of an exactly solved \\ spin-1/2 Ising-Heisenberg orthogonal-dimer chain}

\author[label1]{Lucia G\'alisov\'a\corref{cor1}}
\cortext[cor1]{Corresponding author}
\ead{galisova.lucia@gmail.com}
\author[label2]{Jozef Stre\v{c}ka}
\author[label3]{Taras Verkholyak}
\author[label2]{Samuel Havadej}

\address[label1]{Institute of Manufacturing Management, 
					Faculty of Manufacturing Technologies with the seat in Pre\v{s}ov, Technical University of~Ko\v{s}ice, \\ Bayerova 1, 080\,01 Pre\v{s}ov, Slovakia}
\address[label2]{Institute of Physics, Faculty of Science, 
					 P.~J.~\v{S}af\'arik University, Park Angelinum 9, 040\,01 Ko\v{s}ice, Slovakia}
\address[label3]{Institute for Condensed Matter Physics, National Academy of Sciences of Ukraine, Svientsitskii Street 1, 790\,11 L'viv, Ukraine}

\begin{abstract}
Spin-1/2 orthogonal-dimer chain composed of regularly alternating Ising and Heisenberg dimers is exactly solved in a presence of the magnetic field by the transfer-matrix method. It is shown that the ground-state phase diagram involves in total six different phases. Besides the ferromagnetic phase with fully polarized spins one encounters the singlet antiferromagnetic and modulated antiferromagnetic phases manifested in zero-temperature magnetization curves as zero magnetization plateau, the frustrated ferrimagnetic and singlet ferrimagnetic phases causing existence of an intermediate one-half magnetization plateau, and finally, the intriguing modulated ferrimagnetic phase with a translationally broken symmetry leading to an unconventional one-quarter magnetization plateau. The quantum character of individual ground states is quantified via the concurrence, which measures a strength of the bipartite entanglement within the pure and mixed states of the Heisenberg dimers at zero as well as nonzero temperatures. The parameter region, where the bipartite entanglement may be in contrast to general expectations reinforced upon increasing of temperature and/or magnetic field, is elucidated. 
\end{abstract}



\end{frontmatter}


\section{Introduction}
\label{intro}

The gapped quantum ground states manifested in low-tem\-perature magnetization curves as intermediate plateaus remain at a forefront of intense theoretical studies, because several intriguing fractional magnetization plateaus were experimentally detected in high-field magnetization curves of the magnetic compound SrCu$_2$(BO$_3$)$_2$~\cite{kage99,oniz00,kage01,seba08,mats13} providing a long-sought experimental realization of the Shastry-Sutherland model~\cite{shas81,miya99}. Despite substantial efforts, the number and microscopic nature of intermediate magnetization plateaus of SrCu$_2$(BO$_3$)$_2$ still remain unresolved issue due to extraordinary mathematical difficulties related to a theoretical modeling of the Shastry-Sutherland model at zero~\cite{dori08,neme12,corb14,verk14,folt14,schn16} as well as nonzero~\cite{wess18,wiet19} temperatures.

By contrast, the magnetization process of one-dimensional counterpart of the Shastry-Sutherland model, which is commonly referred to either as the spin-1/2 Heisenberg orthogonal-dimer or dimer-plaquette chain \cite{richt96,ivan97,richt98}, is quite well established nowadays. Except three most pronounced magnetization plateaus at zero, one-quarter and one-half of the saturation magnetization, one additionally encounters an infinite series of smaller fractional magnetization plateaus at rational numbers $n/(2n+2) = 1/4, 1/3, 3/8, \ldots, 1/2$ ranging in between one-quarter and one-half magnetization plateaus~\cite{schulenburg02,sch02}. Unfortunately, a respective magnetic compound that would enable an experimental testing of this peculiar sequence of the fractional magnetization plateaus is not available to date. 

Recent experimental discovery of the polymeric coordination compound [Dy(hfac)$_2$(CH$_3$OH)]$_2$[Cu(dmg)(Hdmg)]$_2$ \cite{ueki2007,okazawa2008}, which will be further referred to as the polymeric chain [Dy$_2$Cu$_2$]$_n$, has afforded a valuable experimental realization of the spin-1/2 Ising-Heisenberg orthogonal-dimer chain with a regular alternation of the highly anisotropic dimeric units of Dy$^{3+}$ magnetic ions (Ising dimers) with the almost isotropic dimeric units of Cu$^{2+}$ magnetic ions. It is supposed that the most dominant coupling in [Dy$_2$Cu$_2$]$_n$ is by far an antiferromagnetic interaction between the nearest-neighbouring Dy$^{3+}$ and Cu$^{2+}$ magnetic ions, whereas the dinuclear entities of Dy$^{3+}$ ions and Cu$^{2+}$ ions are coupled presumably through much weaker ferromagnetic interaction \cite{okazawa2008}. It is worthwhile to remark that the spin-1/2 Ising-Heisenberg orthogonal-dimer chain with a regularly alternating Ising and Heisenberg dimers arranged in an orthogonal fashion can be exactly solved by adapting the approach elaborated in Refs. \cite{paul13,verk13,verh14} for various versions of this intriguing one-dimensional quantum spin model. 

In the present work we will introduce and rigorously solve a spin-1/2 Ising-Heisenberg orthogonal-dimer chain composed of regularly alternating Ising and Heisenberg dimers in a presence of the external magnetic field. Although the investigated quantum spin chain is somewhat oversimplified in that it does not take into account two different exchange pathways between Dy$^{3+}$ and Cu$^{2+}$ magnetic ions existing within the polymeric compound [Dy$_2$Cu$_2$]$_n$, we believe that this quantum spin chain may shed light on the nature of unconventional quantum ground states invoked by the external magnetic field in a low-temperatu\-re magnetization process of the polymeric complex [Dy$_2$Cu$_2$]$_n$.

The organization of this paper is as follows. In Section~\ref{sec:2} we will introduce and solve the spin-1/2 Ising-Heisenberg orthogo\-nal-dimer chain within the framework of the transfer-matrix method.  Section~\ref{sec:3} includes a comprehensive discussion of the most interesting results obtained for the ground-state phase diagram, the magnetization process and the bipartite entanglement emergent within the Heisenberg dimers. The most important findings and future outlooks are briefly mentioned in Section~\ref{sec:4}.

\section{Model and its exact solution}
\label{sec:2}
\begin{figure}[t!]
	\vspace{3mm}
	\hspace{-2.25cm}
	\includegraphics[width=1.5\columnwidth]{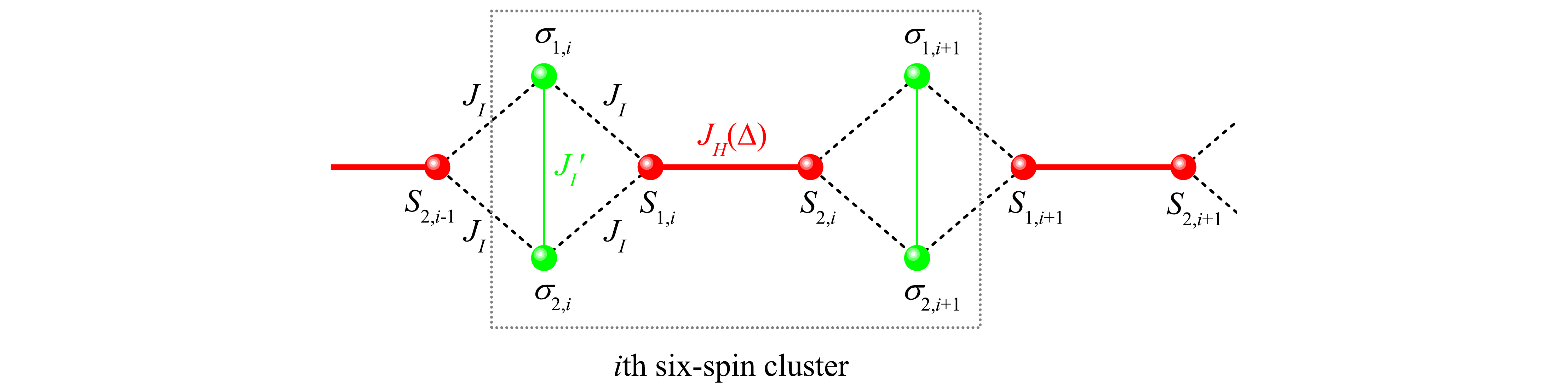}
	\caption{The magnetic structure of the frustrated spin-1/2 Ising-Heisenberg orthogonal-dimer chain. Green (red) balls denote lattice sites occupied by the Ising (Heisenberg) spins, thin green (thick red) lines correspond to the Ising (Heisenberg) intra-dimer bonds and dashed black lines illustrate the Ising inter-dimer bonds. The dotted rectangle represents the $i$th six-spin cluster described by the cluster Hamiltonian~(\ref{eq:Hi}).}
	\label{figure1}
\end{figure}

In the present paper, we will consider the quantum spin-$1/2$ Ising-Heisenberg orthogonal-dimer chain schematically depict\-ed in Fig.~\ref{figure1} and defined through the total Hamiltonian:
\begin{eqnarray}
\label{eq:Htotal}
\hat{H} \!\!\!&=&\!\!\!\!
J_{H} \sum_{i = 1}^{N} \big(\hat{\mathbf S}_{1,i}\cdot\hat{\mathbf S}_{2,i}\big)_{\Delta} + 
J_{I}^{\prime} \sum_{i = 1}^{N} \hat{\sigma}_{1,i}^{z}\hat{\sigma}_{2,i}^{z} \nonumber \\
\!\!\!&+&\!\!\!\!\! J_{I} \sum_{i = 1}^{N} \big[\hat{S}_{1,i}^{z}\big(\hat{\sigma}_{1,i}^{z}\!+\hat{\sigma}_{2,i}^{z}\big) +
\hat{S}_{2,i}^{z}\big(\hat{\sigma}_{1,i+1}^{z}\!+\hat{\sigma}_{2,i+1}^{z}\big)\big] \nonumber \\ 
\!\!\!&-&\!\!\!\!\! h_{H}\sum_{i = 1}^{N} \big(\hat{S}_{1,i}^{z} + \hat{S}_{2,i}^{z}\big)
- h_{I}\sum_{i = 1}^{N} \big(\hat{\sigma}_{1,i}^{z}\!+\hat{\sigma}_{2,i}^{z}\big).
\end{eqnarray}
In above, $\big(\hat{\mathbf S}_{1,i}\cdot\hat{\mathbf S}_{2,i}\big)_{\Delta}=\Delta(\hat{S}_{1,i}^x\hat{S}_{2,i}^x + \hat{S}_{1,i}^y\hat{S}_{2,i}^y)+\hat{S}_{1,i}^z\hat{S}_{2,i}^z$, $\hat{S}_{1(2),i}^{\alpha}$ ($\alpha = x,y,z$) label the spatial components of the standard spin-1/2 operators corresponding to the Heisenberg spins forming horizontal dimers, $\hat{\sigma}_{1(2),i}^z$ are the spatial components of the standard spin-1/2 operators related to the Ising spins forming vertical dimers, the parameter $J_H$ denotes the XXZ Heisenberg intra-dimer interaction on horizontal bonds with the parameter of exchange anisotropy $\Delta$, $J_{I}^{\prime}$ represents the Ising intra-dimer interaction between the Ising spins on vertical bonds, and $J_{I}$ denotes the Ising inter-dimer interaction between the nearest-neighboring Ising and Heisenberg spins. The last two terms $h_H$ and $h_I$ are Zeeman terms, which account for the magnetostatic energy of the Heisenberg and Ising spins in an applied longitudinal magnetic field, respectively. Finally, $N$ denotes the total number of the Heisenberg and Ising spin dimers and the periodic boundary conditions $\hat{\sigma}_{1(2),N+1}^{z} \equiv \hat{\sigma}_{1(2),1}^{z}$ are assumed for the sake of simplicity. 

For further convenience, the total Hamiltonian (\ref{eq:Htotal}) of the spin-$1/2$ Ising-Heisenberg orthogonal-dimer chain can be alternatively rewritten as a sum of the six-spin cluster Hamiltonians schematically delimited in Fig.~\ref{figure1} by a dotted rectangle: 
\begin{eqnarray}
\label{eq:H}
\hat{H} \!\!\!&=&\!\!\!\!
\sum_{i = 1}^{N} \hat{H}_{i},
\\
\label{eq:Hi}
\hat{H}_{i}\!\!\!&=&\!\!\!\!
J_{H}\big(\hat{\mathbf S}_{1,i}\cdot\hat{\mathbf S}_{2,i}\big)_{\Delta}
+\frac{J_{I}^{\prime}}{2}\big(\hat{\sigma}_{1,i}^{z}\hat{\sigma}_{2,i}^{z}+\hat{\sigma}_{1,i+1}^{z}\hat{\sigma}_{2,i+1}^{z}\big) 
\nonumber\\ 
\!\!\!&+&\!\!\!\!\! 
J_{I}\big[
\hat{S}_{1,i}^{z}\big(\hat{\sigma}_{1,i}^{z}\!+\hat{\sigma}_{2,i}^{z}\big) +
\hat{S}_{2,i}^{z}\big(\hat{\sigma}_{1,i+1}^{z}\!+\hat{\sigma}_{2,i+1}^{z}\big)\big]
\nonumber\\
\!\!\!&-&\!\!\!\!\!
h_{H}\big(\hat{S}_{1,i}^{z} + \hat{S}_{2,i}^{z}\big)
-\frac{h_{I}}{2}\big(\hat{\sigma}_{1,i}^{z}\!+\hat{\sigma}_{2,i}^{z}\!+\hat{\sigma}_{1,i+1}^{z}\!+\hat{\sigma}_{2,i+1}^{z}\big).
\end{eqnarray}
It is noteworthy that the $i$th six-spin cluster Hamiltonian (\ref{eq:Hi}) involves the vertical Ising dimers from two adjacent unit cells, whereas the factor $1/2$ emergent at the Ising coupling $J_{I}^{\prime}$ and the Zeeman term $h_I$ avoids a double counting of these two interaction terms being symmetrically split into two consecutive cluster Hamiltonians.  

It is obvious from Eq.~(\ref{eq:Hi}) that different cluster Hamiltonians satisfy the standard commutation relation \mbox{$[\hat{H}_{i}, \hat{H}_{j}]=0$}, and thus, the determination of eigenvalues of $\hat{H}_{i}$ is sufficient to exactly resolve the considered spin-$1/2$ Ising-Heisenberg orthogo\-nal-dimer chain. The relevant calculation can be performed in a matrix representation of the Hilbert subspace spanned over the orthonormal basis of four available spin states corresponding to the $i$th Heisenberg spin pair 
$\big\{|\!\uparrow\uparrow\rangle_{i} = |\!\uparrow\rangle_{1,i}|\!\uparrow\rangle_{2,i},\, |\!\uparrow\downarrow\rangle_{i} =\linebreak  |\!\uparrow\rangle_{1,i}|\!\downarrow\rangle_{2,i},\,
|\!\downarrow\uparrow\rangle_{i} = |\!\downarrow\rangle_{1,i}|\!\uparrow\rangle_{2,i},\, |\!\downarrow\downarrow\rangle_{i} = |\!\downarrow\rangle_{1,i}|\!\downarrow\rangle_{2,i}\big\}$,
where $|\!\uparrow\rangle_{1(2),i}$  and $|\!\downarrow\rangle_{1(2),i}$ denote eigenvectors of the spin operator $\hat{S}_{1(2),i}^{z}$ with the eigenvalues $S_{1(2),i}^{z} = 1/2$ and $-1/2$, respectively. As a result, one obtains the following set of eigenvalues:
\begin{subequations}
\label{eq:eigenvalues_of_H_i}
\begin{flalign}
\label{eq:eigenvalue1}
E_{i,1}=&
\,\frac{J_{H}}{4} +
\frac{J_{I}^{\prime}}{2}\big(\sigma_{1,i}^{z}\sigma_{2,i}^{z}+\sigma_{1,i+1}^{z}\sigma_{2,i+1}^{z}\big) \nonumber\\ 
& 
+\frac{J_{I} - h_{I}}{2}\big(\sigma_{1,i}^{z}+\sigma_{2,i}^{z}+\sigma_{1,i+1}^{z}+\sigma_{2,i+1}^{z}\big)
-
h_{H},
\\
\label{eq:eigenvalue2}
E_{i,2}=&
\,\frac{J_{H}}{4} +
\frac{J_{I}^{\prime}}{2}\big(\sigma_{1,i}^{z}\sigma_{2,i}^{z}+\sigma_{1,i+1}^{z}\sigma_{2,i+1}^{z}\big)
\nonumber\\ 
& 
-\frac{J_{I}+h_{I}}{2}\big(\sigma_{1,i}^{z}+\sigma_{2,i}^{z}+\sigma_{1,i+1}^{z}+\sigma_{2,i+1}^{z}\big)
+
h_{H},
\\
\label{eq:eigenvalue3}
E_{i,3}=&
-\frac{J_{H}}{4}  
+
\frac{J_{I}^{\prime}}{2}\big(\sigma_{1,i}^{z}\sigma_{2,i}^{z}+\sigma_{1,i+1}^{z}\sigma_{2,i+1}^{z}\big)
\nonumber \\
&
+\frac{1}{2}\!\sqrt{J_{I}^{2}\big(\sigma_{1,i}^{z}+\sigma_{2,i}^{z}-\sigma_{1,i+1}^{z}-\sigma_{2,i+1}^{z}\big)^{2}+(J_{H}\Delta)^{2}} 
\nonumber\\ 
& 
-\frac{h_{I}}{2}\big(\sigma_{1,i}^{z}+\sigma_{2,i}^{z}+\sigma_{1,i+1}^{z}+\sigma_{2,i+1}^{z}\big),
\\
\label{eq:eigenvalue4}
E_{i,4}=&
-\frac{J_{H}}{4}  
+
\frac{J_{I}^{\prime}}{2}\big(\sigma_{1,i}^{z}\sigma_{2,i}^{z}+\sigma_{1,i+1}^{z}\sigma_{2,i+1}^{z}\big)
\nonumber\\ 
& 
-\frac{1}{2}\!\sqrt{J_{I}^{2}\big(\sigma_{1,i}^{z}+\sigma_{2,i}^{z}-\sigma_{1,i+1}^{z}-\sigma_{2,i+1}^{z}\big)^{2}+(J_{H}\Delta)^{2}}
\nonumber\\ 
& 
-\frac{h_{I}}{2}\big(\sigma_{1,i}^{z}+\sigma_{2,i}^{z}+\sigma_{1,i+1}^{z}+\sigma_{2,i+1}^{z}\big),
\end{flalign}
\end{subequations}
and corresponding eigenvectors:
\begin{subequations}
\label{eq:eigenvectors_of_H_i}
\begin{flalign}
\label{eq:eigenvector1}
|\psi\rangle_{i,1} &= |\!\uparrow\uparrow\rangle_{i},
\\
\label{eq:eigenvector2}
|\psi\rangle_{i,2} &= |\!\downarrow\downarrow\rangle_{i},
\\
\label{eq:eigenvector3}
|\psi\rangle_{i,3} &= \sin\varphi_{i}\,|\!\uparrow\downarrow\rangle_{i} + \cos\varphi_{i}\,|\!\downarrow\uparrow\rangle_{i},  
\\
\label{eq:eigenvector4}
|\psi\rangle_{i,4} &= \sin\varphi_{i}\,|\!\uparrow\downarrow\rangle_{i} - \cos\varphi_{i}\,|\!\downarrow\uparrow\rangle_{i},
\end{flalign}
\end{subequations}
where $\tan\left(2\varphi_{i}\right) = J_{H}\Delta/\big[J_{I}\big(\sigma_{1,i}^{z}+\sigma_{2,i}^{z}-\sigma_{1,i+1}^{z}-\sigma_{2,i+1}^{z}\big)\big]$.

Having the full set of eigenvalues of the cluster Hamiltonian~(\ref{eq:Hi}), the partition function $Z$ of the investigated quantum spin chain can be derived by applying the standard transfer-matrix approach~\cite{Bax82,Str15}: 
\begin{eqnarray}
\label{eq:Z}
Z \!\!\!&=&\!\!\!\!\!\!\! \sum_{\{\sigma_{1,i},\sigma_{2,i}\}}\prod_{i = 1}^N{\rm Tr}_{S_{1,i},S_{2,i}}{\rm exp}\big(\!-\beta H_i\big) 
=\!\!\! \sum_{\{\sigma_{1,i},\sigma_{2,i}\}}\prod_{i = 1}^N\sum_{j=1}^{4}{\rm exp}\big(\!-\beta E_{i,j}\big) 
\nonumber\\
&=&\!\!\!\!\!\! \sum_{\{\sigma_{1,i},\sigma_{2,i}\}}\prod_{i = 1}^N {\rm T}\big(\sigma_{1,i}^{z},\sigma_{2,i}^{z};\sigma_{1,i+1}^{z},\sigma_{2,i+1}^{z}\big) = {\rm Tr\,{\bf T}}^N = \sum_{j=1}^{4}\lambda_{j}^{N}.
\end{eqnarray}
In above, $\beta = 1/(k_{\rm B}T)$ is the inverse temperature ($k_{\rm B}$ is the Boltzmann's constant and $T$ is the absolute temperature), the symbol $\sum_{\{\sigma_{1,i},\sigma_{2,i}\}}$ denotes a summation over all possible configurations of the Ising spins from all vertical bonds, the product $\prod_{i = 1}^N$ runs over all six-spin clusters visualized in Fig.~\ref{figure1} and ${\rm Tr}_{S_{1,i},S_{2,i}}$ stands for a trace over degrees of freedom of the $i$th Heisenberg spin dimer. Apparently, the applied formalism enables one to express the partition function $Z$ of the spin-1/2 Ising-Heisenberg orthogonal-dimer chain in terms of four eigenvalues $\lambda_1$, $\lambda_2$, $\lambda_3$, $\lambda_4$ of the $4\times4$ transfer matrix ${\rm\bf T}$, whose elements are formed by 16 Boltzmann's weights corresponding to all available states of two adjacent Ising spin dimers from the $i$th six-spin cluster (see Fig. \ref{figure1}) as defined by the formula:   
\begin{eqnarray}
\label{eq:T}
\lefteqn{{\rm T}\big(\sigma_{1,i}^{z},\sigma_{2,i}^{z};\sigma_{1,i+1}^{z},\sigma_{2,i+1}^{z}\big) =
2{\rm exp}\left[-\frac{\beta J_{I}^{\prime}}{2}\big(\sigma_{1,i}^{z}\sigma_{2,i}^{z}+\sigma_{1,i+1}^{z}\sigma_{2,i+1}^{z}\big)\right]}
\nonumber\\
\!\!\!\!\!\!\! &\times&\!\!\!\!\! {\rm exp}\left[\frac{\beta J_H}{4}+\frac{\beta h_{I}}{2}\big(\sigma_{1,i}^{z}+\sigma_{2,i}^{z}+\sigma_{1,i+1}^{z}+\sigma_{2,i+1}^{z}\big)\right]
\nonumber\\
\!\!\!\!\!\!\! 
&\times&\!\!\!\!\!  
\Bigg\{
{\rm exp}\left(-\frac{\beta J_H}{2}\right)
\cosh\left[\frac{\beta J_{I}}{2}\big(\sigma_{1,i}^{z}+\sigma_{2,i}^{z}+\sigma_{1,i+1}^{z}+\sigma_{2,i+1}^{z}\big) - \beta h_{H}\right] 
\nonumber\\
\!\!\!\!\!\!\! 
&+&\!\!\!\!\! 
\cosh\left[\frac{\beta}{2}\!
\sqrt{J_{I}^{2}\big(\sigma_{1,i}^{z}+\sigma_{2,i}^{z}-\sigma_{1,i+1}^{z}-\sigma_{2,i+1}^{z}\big)^{2} + (J_{H}\Delta)^{2}}\right]
\Bigg\}.
\end{eqnarray}
From the physical point of view, the transfer matrix (\ref{eq:T}) represents the effective Boltzmann's factor, which was obtained after tracing out spin degrees of freedom of two Heisenberg spins from the $i$th horizontal dimer.
Explicit expressions of the transfer-matrix eigenvalues emerging in the final form of the partition function~(\ref{eq:Z}) are:
\begin{subequations}
\begin{flalign}
\label{eq:lambda_1}	
\lambda_{1} &= 0,\hspace{1cm}\\
\label{eq:lambda_2-4}	
\lambda_{j} &= \frac{a}{3} + \frac{2}{3}{\rm sgn}(q)\sqrt{p}\cos\left[\frac{1}{3}\tan^{-1}\left(\frac{\sqrt{p^{3}-q^{2}}}{q}\right) + \frac{2\pi (j-2)}{3}\right]\nonumber\\
&&\hspace{-3.5cm} (j=2,3,4),\,\hspace{0.75cm} 
\end{flalign}
\end{subequations}
where:
\begin{eqnarray}
a\!\!\!&=&\!\!\!A_{1} + 2A_{0} + A_{-1}, 
\nonumber\\
p\!\!\!&=&\!\!\!a^2 - 3(A_{1}A_{-1} + 2A_{0}A_{-1} + 2A_{0}A_{1} - 2B_{1}^2 - 2B_{-1}^2 - B_{0}^2), 
\nonumber
\end{eqnarray}
\begin{eqnarray}
q\!\!\!&=&\!\!\!a^3 - 9a(A_{1}A_{-1} + 2A_{0}A_{-1} + 2A_{0}A_{1} - 2B_{1}^2 - 2B_{-1}^2 - B_{0}^2) {}
\nonumber\\ && 
+ 27(A_{1}A_{0}A_{-1} - A_{-1}B_{1}^2 - A_{1}B_{-1}^2 - A_{0}B_{0}^2 + 2B_{1}B_{0}B_{-1}).
\nonumber
\end{eqnarray}
The coefficients $A_x$ and $B_x$ ($x = -1,0,1$) entering into the formula~(\ref{eq:lambda_2-4}) either directly, or through the parameters $p$, $q$, are given by:
\begin{eqnarray}
A_{x} \!\!\!&=&\!\!\! 2{\rm exp}\left[\frac{\beta J_{H}}{4} + \frac{\beta J_{I}^{\prime}(-1)^{x}}{4} + \beta h_{I}x\right]
\nonumber \\
 \!\!\!& &\!\!\!\!\times
\left[2
{\rm exp}\left(-\frac{\beta J_{H}}{2}\right)\cosh\left(\beta J_{I}x - \beta h_{H}\right) + \cosh\left(\frac{\beta J_{H}\Delta}{2}\right)
\right],
\nonumber\\
B_{x} \!\!\!&=&\!\!\! 2{\rm exp}\left[\frac{\beta J_{H}}{4} + \frac{\beta J_{I}^{\prime}(x^2-1)}{4} + \frac{\beta h_{I}x}{2}\right]
\nonumber \\
\!\!\!& &\!\!\!\!\times\,
\Bigg\{
{\rm exp}\left(-\frac{\beta J_{H}}{2}\right)\cosh\left(\frac{\beta J_{I}x}{2} - \beta h_{H}\right) 
\nonumber \\
\!\!\!& &\hspace{0.3cm}+
\cosh\left[\frac{\beta}{2}\!\sqrt{J_{I}^2(x^2-2)^2 + (J_{H}\Delta)^2}\right]
\Bigg\}.
\nonumber 
\end{eqnarray}
After the explicit form of the transfer-matrix eigenvalues (\ref{eq:lambda_1})-(\ref{eq:lambda_2-4}) is substituted into the final expression for the partition function (\ref{eq:Z}) one obtains a crucial result of our calculations, from which the whole thermodynamics of the spin-1/2 Ising-Heisenberg orthogonal-dimer chain directly follows.
As a matter of fact, the Gibbs free energy $G$ per elementary unit cell can be expressed in the thermodynamic limit $N\to\infty$ in terms of the largest transfer-matrix eigenvalue:
\begin{eqnarray}
G = -k_{\rm B}T\lim_{N\to\infty}\frac{1}{N}\ln Z = -k_{\rm B}T\ln \left(\max\{\lambda_0, \lambda_1,\lambda_2,\lambda_3\}\right).
\end{eqnarray}
Other important physical quantities, such as the local magnetization $m_I = \langle\hat{\sigma}_{1,i}^{z}+\hat{\sigma}_{2,i}^{z}\rangle/2$, $m_H = \langle\hat{S}_{1,i}^{z}+\hat{S}_{2,i}^{z}\rangle/2$ per Ising and Heisenberg spin, respectively, the total magnetization $m$ per lattice site, as well as the pair correlation functions $c_{II}^{z} = \langle\hat{\sigma}_{1,i}^{z}\hat{\sigma}_{2,i}^{z}\rangle$, $c_{HH}^{xx(yy)} = \langle\hat{S}_{1,i}^{x}\hat{S}_{2,i}^{x}\rangle = \langle\hat{S}_{1,i}^{y}\hat{S}_{2,i}^{y}\rangle$, $c_{HH}^{zz} = \langle\hat{S}_{1,i}^{z}\hat{S}_{2,i}^{z}\rangle$ and 
$c_{IH}^{zz} = \langle\hat{S}_{1,i}^{z}\hat{\sigma}_{1,i}^{z}\rangle=\langle\hat{S}_{1,i}^{z}\hat{\sigma}_{2,i}^{z}\rangle
=\langle\hat{S}_{2,i}^{z}\hat{\sigma}_{1,i+1}^{z}\rangle=\langle\hat{S}_{2,i}^{z}\hat{\sigma}_{2,i+1}^{z}\rangle$, 
which bring insight into the local order of the nearest-neighbour spins can be subsequently obtained by means of the differential calculus:
\begin{eqnarray}
\label{eq:mag}
m_I \!\!\!&=&\!\!\! -\frac{1}{2}\frac{\partial G}{\partial h_I},\quad
m_H = -\frac{1}{2}\frac{\partial G}{\partial h_H},\quad
m = \frac{1}{2}\left(m_I + m_H\right),\hspace{0.75cm}
\\
\label{eq:cor1}
c_{II}^{zz} \!\!\!&=&\!\!\! \frac{\partial G}{\partial J_{I}^{\prime}},\hspace{1.5cm}
c_{HH}^{xx(yy)} =  \frac{1}{2J_{H}}\frac{\partial G}{\partial \Delta}, 
\\
\label{eq:cor2}
c_{HH}^{zz}  \!\!\!&=&\!\!\! \frac{\partial G}{\partial J_{H}} -\frac{\Delta}{J_{H}}\frac{\partial G}{\partial \Delta}, 
\quad
c_{IH}^{zz} = \frac{1}{4}\frac{\partial G}{\partial J_{I}}.
\end{eqnarray}
The knowledge of rigorous results for the local magnetization $m_H$ and pair correlation functions $c_{HH}^{xx(yy)}$ and $c_{HH}^{zz}$ corresponding to the Heisenberg spin pairs gives the opportunity to rigorously calculate an interesting physical quantity called concurrence according to the formula~\cite{Woo98, Ami04, Ami08, Ost13}:
\begin{eqnarray}
\label{eq:C}
C = \max\left\{
0, 4|c_{HH}^{xx(yy)}|-2\!\sqrt{\bigg(\frac{1}{4}+c_{HH}^{zz}\bigg)^2 - m_{H}^{2}}\,\right\}\!.
\end{eqnarray}
The quantity~(\ref{eq:C}) represents a feasible measure of bipartite \linebreak entanglement of the Heisenberg spins forming the horizontal dimers at zero as well as nonzero temperatures.

\section{Results and discussion}
\label{sec:3}

In this section, we will discuss a diversity of zero-tempera\-ture spin arrangements, magnetization process and bipartite entanglement of the particular version of the quantum spin-$1/2$ Ising-Heisenberg orthogonal-dimer chain with the antiferromagnetic exchange interactions $J_{H}>0$, $J_{I}>0$ and $J_{I}^{\prime}>0$. Without loss of generality, we will restrict our analysis to the special case of the orthogonal-dimer chain with the isotropic Heisenberg intra-dimer interaction (the case $\Delta=1$), which exhibits all generic features of the more general quantum spin-$1/2$ Ising-Heisenberg orthogonal-dimer chain with the anisotropic XXZ Heisenberg intra-dimer interaction with $\Delta\neq1$. To reduce a number of free parameters, we will also assume the same magnetic fields acting on the Ising and Heisenberg spins $h_I = h_H = h$, which corresponds to setting the same $g$-factors for these spins from the physical point of view.

\subsection{Ground-state phase diagram}
\label{subsec:3.1}

Let us start by analyzing possible zero-temperature spin arrangements of the considered quantum spin chain. By comparing the eigenvalues~(\ref{eq:eigenvalue1})--(\ref{eq:eigenvalue4}) for all available configurations of the Ising spins from $i$th and $(i+1)$st vertical dimers one can identify in total six different ground states specified below:
\begin{itemize}
	\item[(i)] The {\it ferromagnetic} (FM) phase -- the unique classical phase with the perfect ferromagnetic arrangement of all the Ising and Heisenberg spins:
	\begin{eqnarray}
	\label{eq:FM}
	|\textrm{FM}\rangle = \prod_{i=1}^{N}
	\left.\big|\mbox{\normalsize${\uparrow\atop\uparrow}$}\right\rangle_{\!i}\otimes
	|\!\uparrow\uparrow\rangle_{i}\,.
	\end{eqnarray}	
	The energy  is: $\displaystyle E_{\textrm{FM}} = \frac{N}{4}\left(J_{H}+ J_{I}^{\prime} + 4J_{I} - 8h\right);$
\end{itemize}
\begin{itemize}
	\item[(ii)] The {\it frustrated ferrimagnetic} (FI) phase -- the macroscopically degenerate ($2^N$) ferrimagnetic phase with the antiferromagnetic spin arrangement on the vertical Ising di\-mers and the ferromagnetic spin arrangement on the horizontal Heisenberg dimers:
	\begin{eqnarray}
	\label{eq:FI}
	|\textrm{FI}\rangle = \prod_{i=1}^{N}
	\left.\big|\mbox{\normalsize${\uparrow\atop\downarrow}$}\right\rangle_{\!i} \,\left(\textrm{or}\left.\big|\mbox{\normalsize${\downarrow\atop\uparrow}$}\right\rangle_{\!i}\right)\otimes
	|\!\uparrow\uparrow\rangle_{i}\,.
	\end{eqnarray}	
	The energy  is: $\displaystyle E_{\textrm{FI}} = \frac{N}{4}\left(J_{H} - J_{I}^{\prime} - 4h\right);$
\end{itemize}
\begin{itemize}
	\item[(iii)] The {\it singlet ferrimagnetic} (SFI) phase --  the unique quantum phase with the ferromagnetic spin arrangement of the vertical Ising dimers and the fully entangled singlet state of the horizontal Heisenberg dimers:
	\begin{eqnarray}	
	\label{eq:SFI}
	|\textrm{SFI}\rangle = \prod_{i=1}^{N}
	\left.\big|\mbox{\normalsize${\uparrow\atop\uparrow}$}\right\rangle_{\!i}\otimes
	\frac{1}{\sqrt{2}}\,\Big(|\!\uparrow\downarrow\rangle_{i} - |\!\downarrow\uparrow\rangle_{i}\Big). 
	\end{eqnarray}
	The energy is: $\displaystyle E_{\textrm{SFI}} = -\frac{N}{4}\left(3J_{H} - J_{I}^{\prime} + 4h\right);$
\end{itemize}
\begin{itemize}
	\item[(iv)] The {\it singlet antiferromagnetic} (SAF) phase -- the macroscopically degenerate ($2^N$) antiferromagnetic phase with the perfect antiferromagnetic spin arrangement of the vertical Ising dimers and the fully entangled singlet state of the horizontal Heisenberg dimers:
	\begin{eqnarray}
	\label{eq:SAF}
	|\textrm{SAF}\rangle = \prod_{i=1}^{N}\left.\big|\mbox{\normalsize${\uparrow\atop\downarrow}$}\right\rangle_{\!i} \,\left(\textrm{or}\left.\big|\mbox{\normalsize${\downarrow\atop\uparrow}$}\right\rangle_{\!i}\right)\otimes
	\frac{1}{\sqrt{2}}\,
	\Big(
	|\!\uparrow\downarrow\rangle_{i} - |\!\downarrow\uparrow\rangle_{i}
	\Big).
	\end{eqnarray}
	The energy is: $\displaystyle E_{\textrm{SAF}} = -\frac{N}{4}\left(3J_{H} + J_{I}^{\prime}\right);$ 
\end{itemize}
\begin{itemize}
	\item[(v)] The {\it modulated ferrimagnetic} (MFI) phase -- the macroscopically degenerate ($2^{N/2}$) phase characterized by a regular alternation of the ferromagnetically and antiferromagnetically ordered vertical Ising dimers and the sing-let-like state of the horizontal Heisenberg dimers:
	\begin{eqnarray}	
	\label{eq:MFI}
	|\textrm{MFI}\rangle = \prod_{i=1}^{N/2}
	\left.\big|\mbox{\normalsize${\uparrow\atop\uparrow}$}\right\rangle_{\!2i-1}\!\otimes\Big(
	\sin\varphi_{1}|\!\uparrow\downarrow\rangle_{2i-1} - \cos\varphi_{1}|\!\downarrow\uparrow\rangle_{2i-1}
	\Big)
	\nonumber\\
	\otimes\!
	\left.\big|\mbox{\normalsize${\uparrow\atop\downarrow}$}\right\rangle_{\!2i} 
	\,\left(\textrm{or}\left.\big|\mbox{\normalsize${\downarrow\atop\uparrow}$}\right\rangle_{\!2i}\right)\otimes\Big(
	\cos\varphi_{1}|\!\uparrow\downarrow\rangle_{2i} - \sin\varphi_{1}|\!\downarrow\uparrow\rangle_{2i}
	\Big)\,.
	\end{eqnarray}
	The energy is: $\displaystyle E_{\textrm{MFI}} = -\frac{N}{4}\left(J_{H} + 2\!\sqrt{J_{I}^{2} +J_{H}^{2}} + 2h\right);$
\end{itemize}
\begin{itemize}
	\item[(vi)] The {\it modulated antiferromagnetic} (MAF) phase -- the two-fold degenerate phase characterized by a regular alternation of two kinds of fully polarized vertical Ising dimers and the other singlet-like state of the horizontal Heisenberg dimers:
	\begin{eqnarray}	
	\label{eq:MAF}
	|\textrm{MAF}\rangle \!\!\!\!&=&\!\!\!\! \prod_{i=1}^{N/2}
	\left.\big|\mbox{\normalsize${\uparrow\atop\uparrow}$}\right\rangle_{\!2i-1}\!\otimes\Big(
	\sin\varphi_{2}|\!\uparrow\downarrow\rangle_{2i-1} - \cos\varphi_{2}|\!\downarrow\uparrow\rangle_{2i-1}
	\Big)\,\,\,
	\nonumber\\
	\!\!\!\!&&\,\,\,\,\otimes\!
	\left.\big|\mbox{\normalsize${\downarrow\atop\downarrow}$}\right\rangle_{\!2i} \otimes\Big(
	\cos\varphi_{2}|\!\uparrow\downarrow\rangle_{2i} - \sin\varphi_{2}|\!\downarrow\uparrow\rangle_{2i}
	\Big).
	\end{eqnarray}
	The energy is: $\displaystyle E_{\textrm{MAF}} = -\frac{N}{4}\left(J_{H} + 2\!\sqrt{4J_{I}^{2} +J_{H}^{2}} - J_{I}^{\prime}\right).$ 
\end{itemize}
Note that the two-site ket vectors with vertically written arrows in the eigenvectors~(\ref{eq:FM})--(\ref{eq:MAF}) determine spin arrangements within the vertical Ising dimers, while the ones with horizontally written arrows determine spin arrangements within the horizontal Heisenberg dimers. Up (down) arrow refers to the spin state $1/2$ ($-1/2$) in both kinds of ket vectors. The mixing angles $\varphi_{1}$ and $\varphi_{2}$ in the last two eigenvectors~(\ref{eq:MFI}) and~(\ref{eq:MAF}), which determine a degree of quantum entanglement within the horizontal Heisenberg spin dimers in the MAF and MFI phases, respectively, are given by the relation $\tan\left(2\varphi_{n}\right) = J_{H}/(nJ_{I})$ ($n=1,\,2$).

\begin{figure*}[t!]
	\centering
	\vspace{-0.5cm}
	\includegraphics[width=1.0\textwidth]{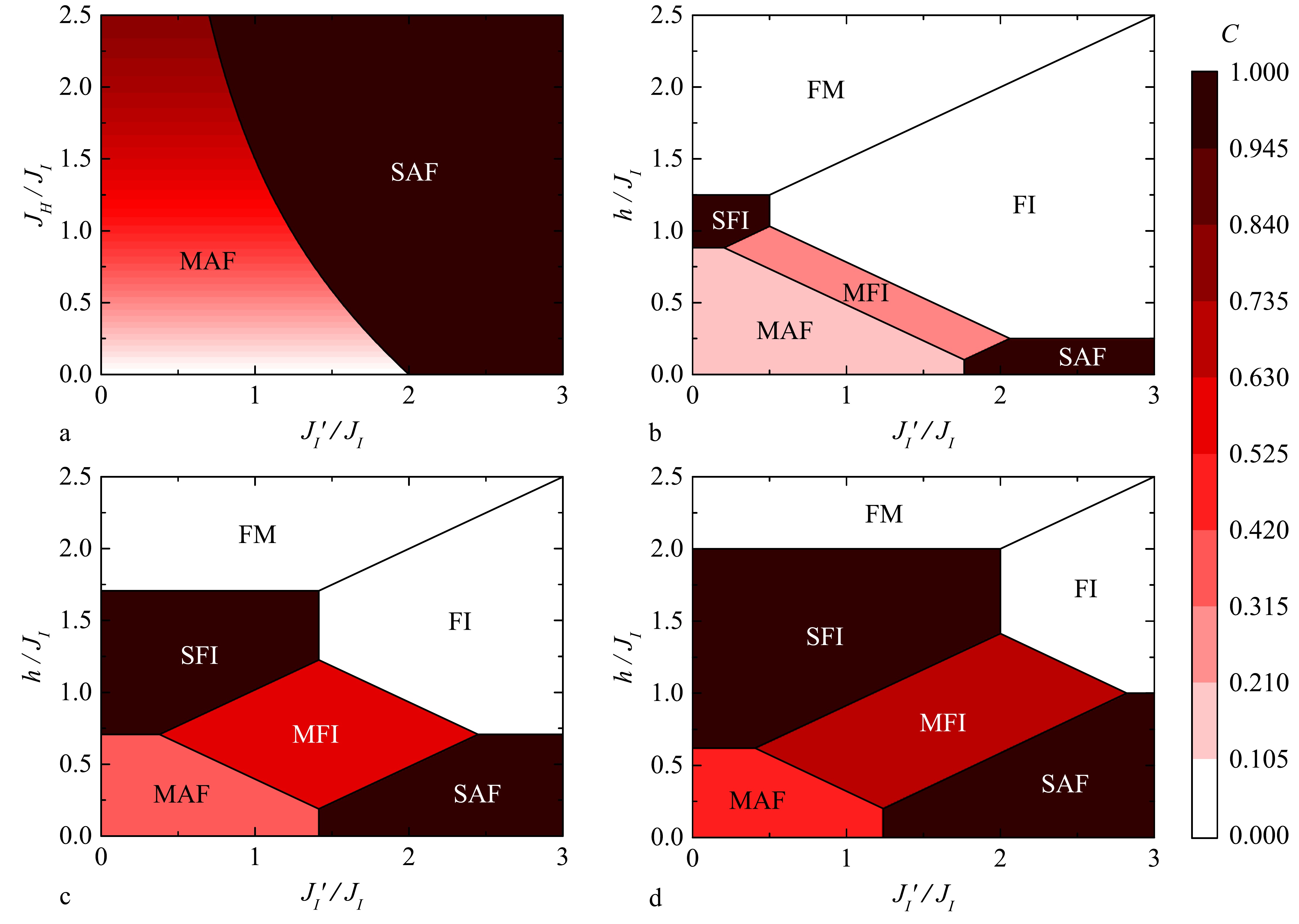}
	\vspace{-0.5cm}
	\caption{The ground-state phase diagrams of the spin-$1/2$ Ising-Heisenberg orthogonal-dimer chain in the $J_{I}^{\prime}/J_{I}-J_{H}/J_{I}$ parameter plane by assuming zero magnetic field $h/J_{I} = 0$ (panel a) and in the $J_{I}^{\prime}/J_{I}-h/J_{I}$ parameter plane for three representative values of the interaction ratio $J_H/J_I=0.25$ (panel b), $J_H/J_I=\sqrt{2}/2$ (panel c) and $J_H/J_I=1$ (panel d). The figures are supplemented with a density plot of the concurrence~$C$ measuring a bipartite entanglement within the horizontal Heisenberg dimers.}
	\label{fig:2}
	\vspace{0.0cm}
\end{figure*}
The overall ground-state behavior of the investigated spin-$1/2$ Ising-Heisenberg orthogonal-dimer chain is depicted in\linebreak Fig.~\ref{fig:2} in the parameter planes $J_{I}^{\prime}/J_{I}-J_{H}/J_{I}$ for $h/J_{I}=0$ (panel~a) and $J_{I}^{\prime}/J_{I}-h/J_{I}$ for three representative values of the interaction ratio $J_{H}/J_{I}= 1/4$, $\sqrt{2}/2$, $1$ (panels~b-d). Black solid lines in the displayed phase diagrams indicate first-order (discontinuous) phase transitions between the coexisting phases. They were analytically calculated by comparing the ground-state energies corresponding to the eigenvectors listed in \linebreak Eqs.~(\ref{eq:FM})--(\ref{eq:MAF}). As one can see from Fig.~\ref{fig:2}a, only two quantum phases SAF and MAF emerge as possible ground states at zero magnetic field $h/J_{I}=0$. The SAF phase is stable in the parameter region $J_{I}^{\prime}/J_{I} > \sqrt{4 + (J_{H}/J_{I})^{2}} - J_{H}/J_{I}$, where the predominant intra-dimer coupling $J_I^{\prime}$ maintains the antiparallel spin alignment of the vertical Ising dimers and the intra-dimer interaction $J_{H}$ is strong enough to create the fully entangled singlet-dimer state on all horizontal Heisenberg dimers. In the rest of the parameter space, the peculiar MAF phase with a two-fold broken translational symmetry due to a regular alternation of two kinds of fully polarized vertical Ising dimers and two alternating kinds of singlet-like states of the horizontal Heisenberg dimers appears. It is noteworthy that the total magnetization is zero within both zero-field ground states SAF and MAF. 

The situation becomes more complex after turning on the external magnetic field. Besides the SAF and MAF phases, three ferrimagnetic phases SFI, MFI and FI with nonzero magnetization can be observed in addition to the fully polarized FM phase due to a mutual interplay between the applied magnetic field $h$ and the coupling constants  $J_{H}$, $J_{I}^{\prime}$, $J_{I}$. It is obvious from Figs.~\ref{fig:2}b-d that the parameter regions corresponding to the quantum phases SAF and SFI (MAF) are gradually extended (reduced) upon increasing value of the interaction ratio $J_{H}/J_{I}$ promoting existence of the singlet-dimer state on the horizontal bonds, while the classical FI and FM phases are contrarily shifted towards higher magnetic fields. As a result, both spin arrangements inherent to  SAF and SFI phases simultaneously appear in a zero-temperature magnetization process for moderate values of the interaction ratio $J_{I}^{\prime}/J_ {I}\in\big(\!\sqrt{4 + (J_{H}/J_{I})^{2}} - J_{H}/J_{I}, 2J_{H}/J_{I}\big)$ after a relative strength of the Heisenberg intra-dimer coupling exceeds the value $J_{H}/J_{I} =\sqrt{2}/2$ (see Fig.~\ref{fig:2}d). Last but not least, the evolution of the parallelogram-shaped parameter space corresponding to the MFI phase, which emerges at moderate values of the interaction ratio $J_{I}^{\prime}/J_{I}$ and the magnetic field $h/J_{I}$ faithfully follows the trend of adjacent phases MAF, SFI, SAF, and FI: the phase boundaries MFI--SFI and SAF--MFI are gradually prolonged, while the ones MAF--MFI, MFI--FI are gradually shortened upon increasing of the interaction ratio $J_{H}/J_{I}$. For the particular value $J_{H}/J_{I} =\sqrt{2}/2$, the parameter region corresponding to the MFI phase has the shape of a rhombus with the shorter diagonal parallel to the field-axis (see Fig.~\ref{fig:2}c).

\subsection{Magnetization process at zero and nonzero temperatures}
\label{subsec:3.2}

\begin{figure*}[ht!]
	\centering
	\vspace{-1.5cm}
	\includegraphics[width=0.45\textwidth]{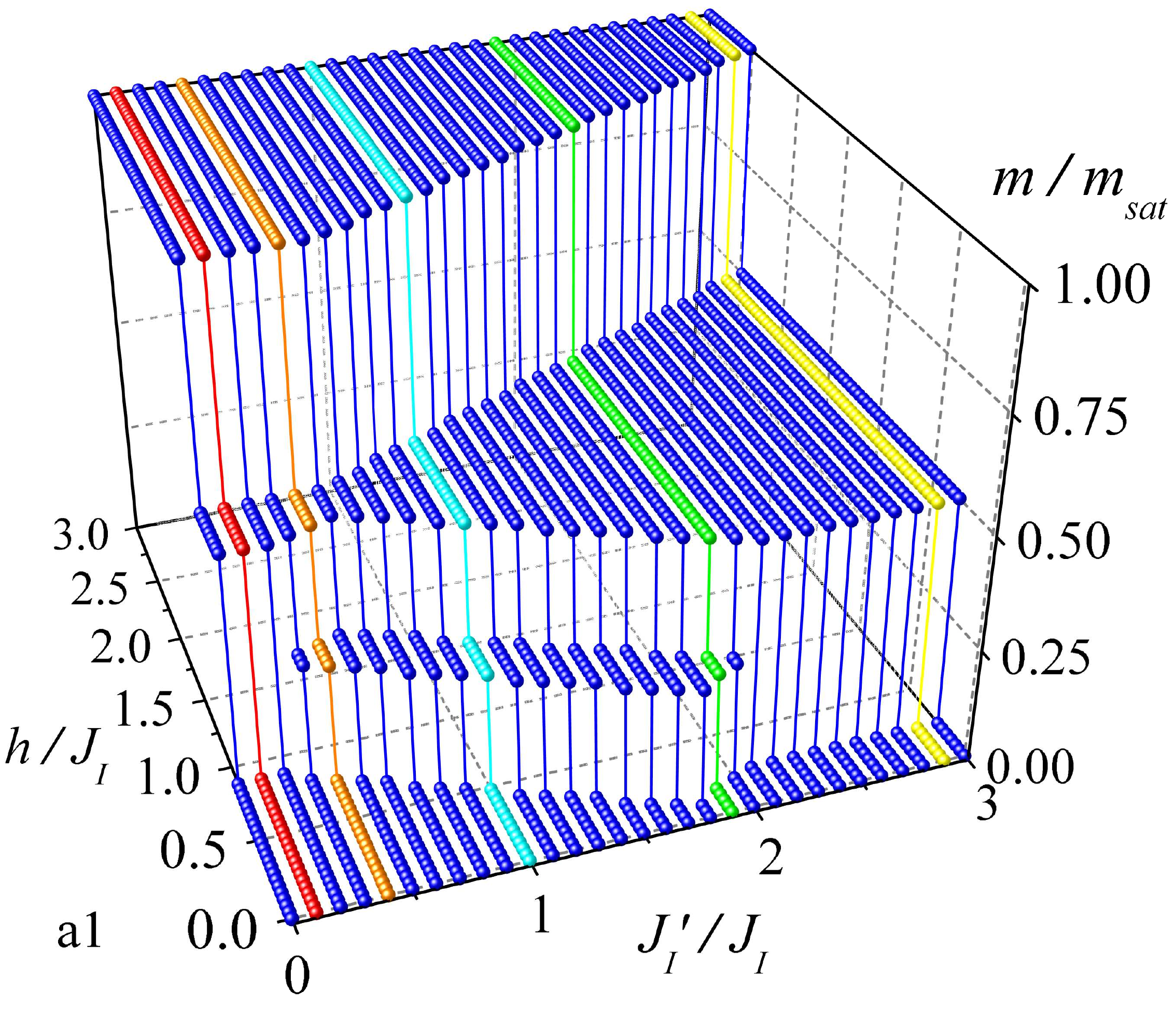}
	\includegraphics[width=0.45\textwidth]{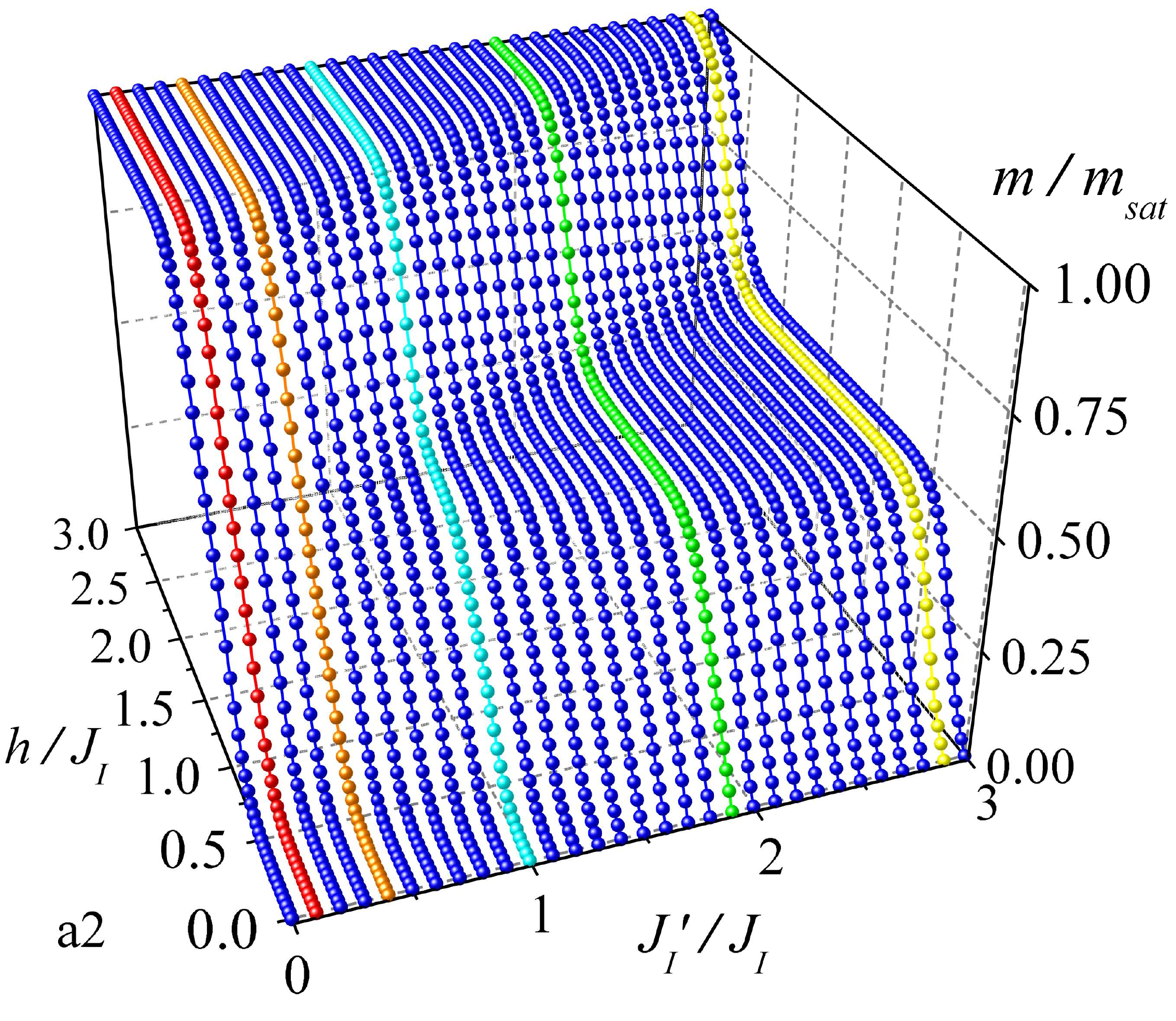}
	\\[0mm]
	\includegraphics[width=0.45\textwidth]{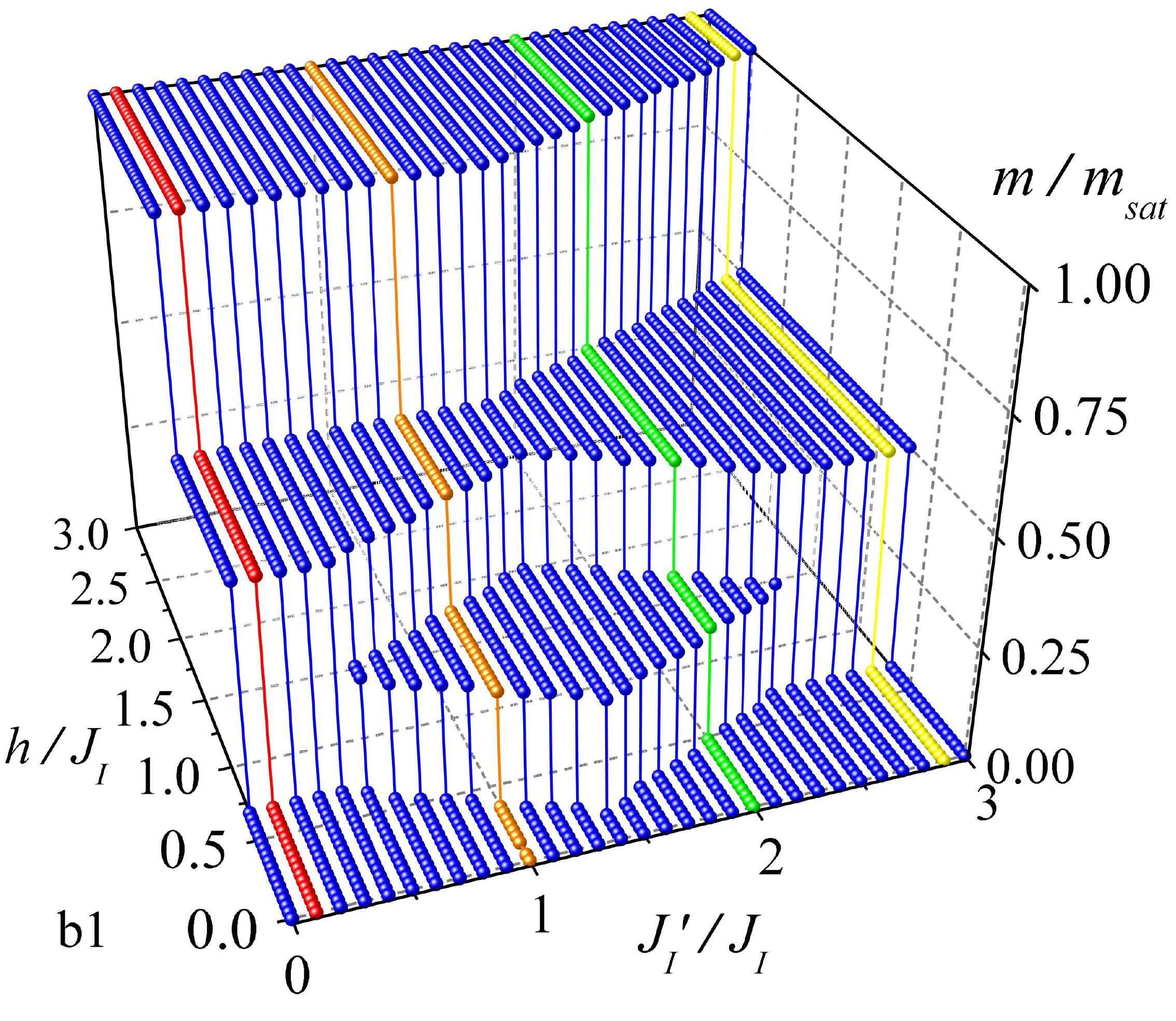}
	\includegraphics[width=0.45\textwidth]{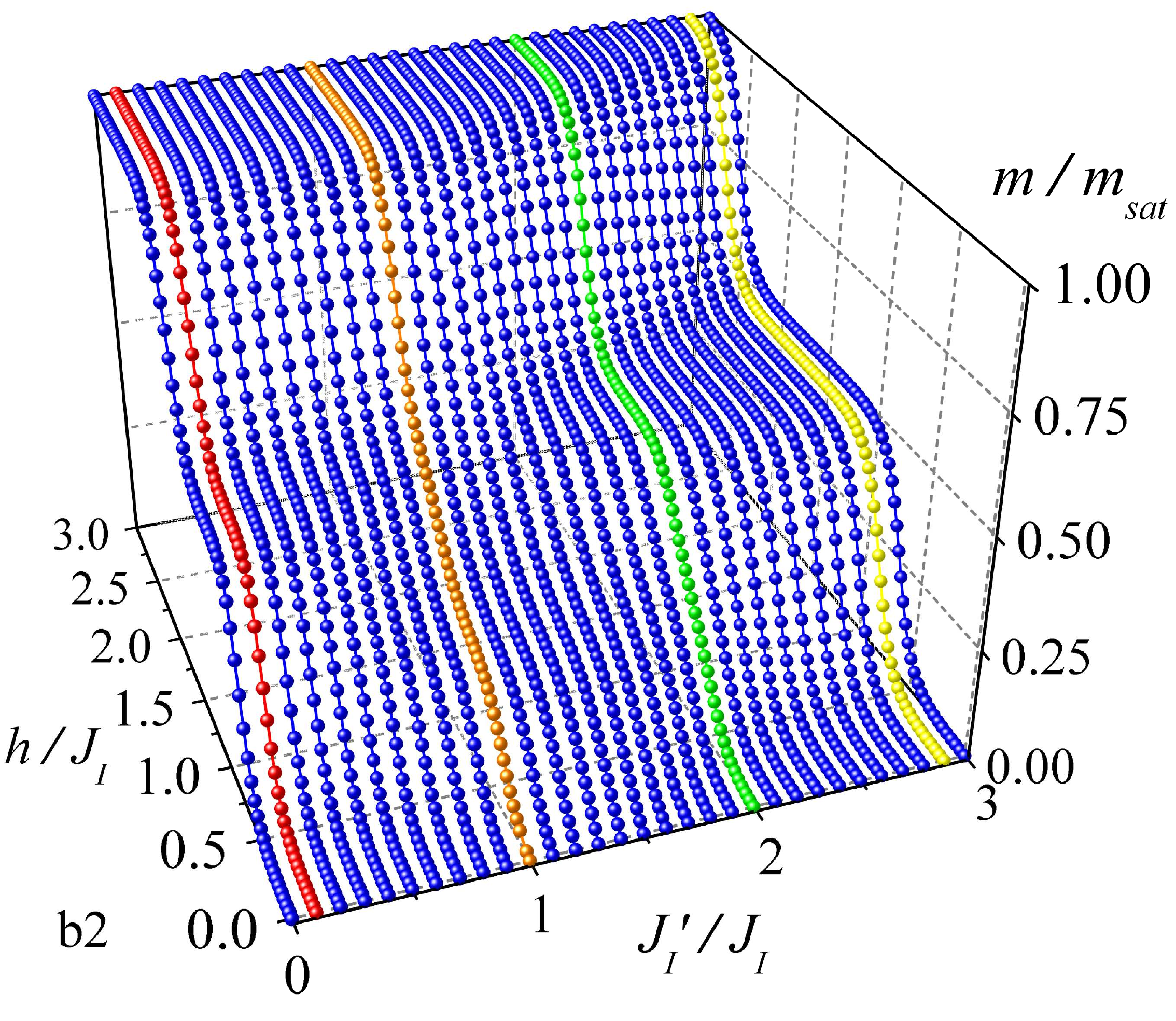}
	\\[0mm]
	\includegraphics[width=0.45\textwidth]{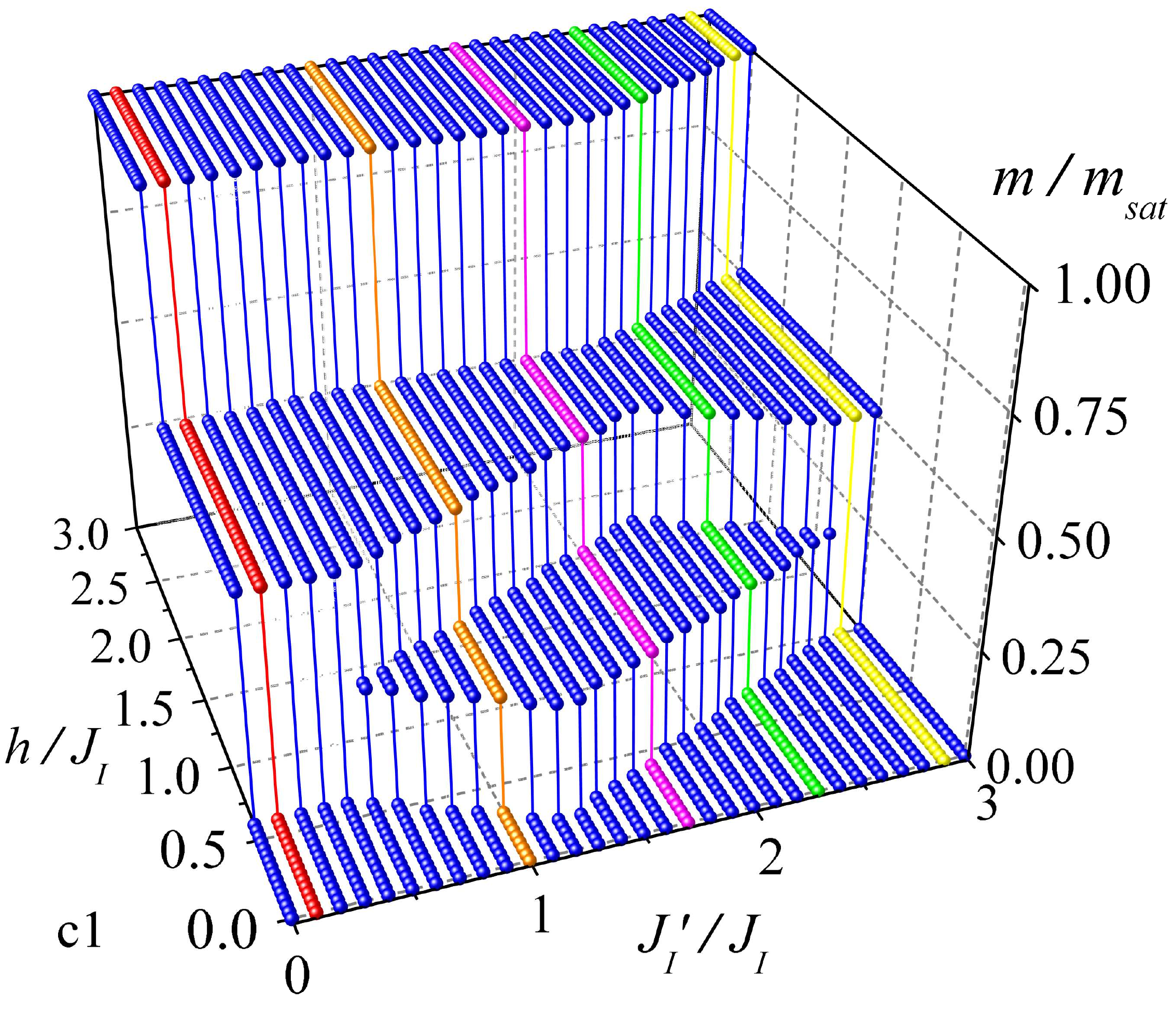}
	\includegraphics[width=0.45\textwidth]{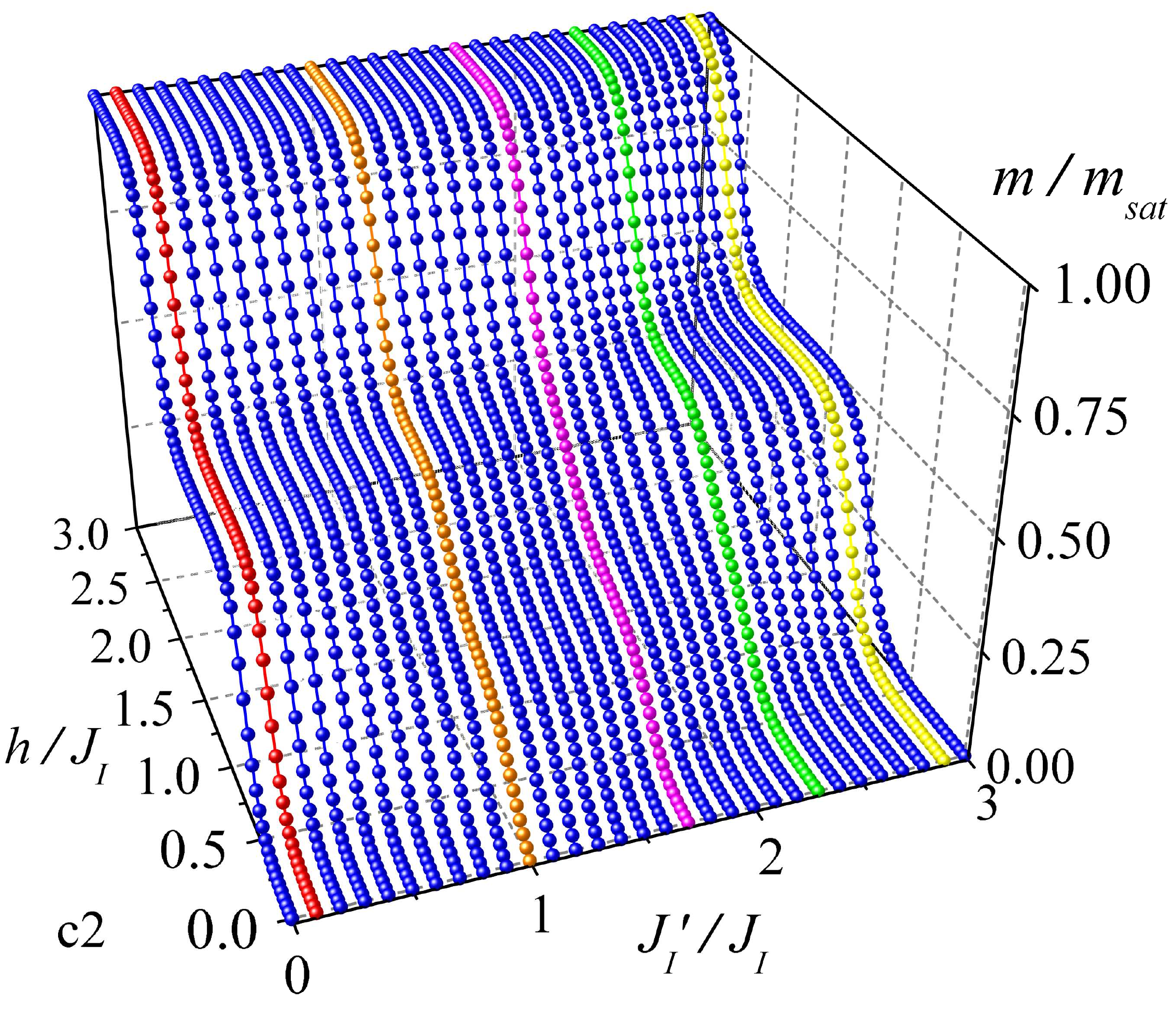}
	\\[2mm]
	\includegraphics[width=0.95\textwidth]{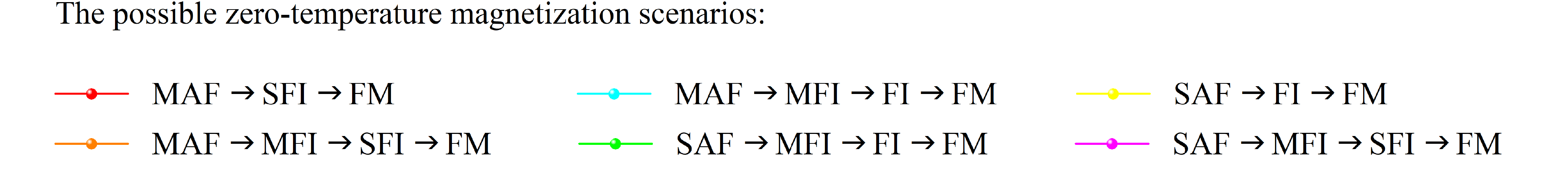}
	\vspace{-0.25cm}
	\caption{3D plots of the total magnetization $m$ reduced with respect to its saturation value $m_{sat}$ as a function of the magnetic field $h/J_{I}$ and the interaction ratio $J_{I}^{\prime}/J_{I}$ for three fixed values of the interaction ratio $J_H/J_I=1/4$ (panels a1, a2), $J_H/J_I=\sqrt{2}/2$ (panels b1, b2) and $J_H/J_I=1$ (panels c1, c2) at two different temperatures $k_{\rm B}T/J_{I} = 0$ (left panels) and $k_{\rm B}T/J_{I} = 0.1$ (right panels). The curves of distinct colors refer to different magnetization scenarios, which can be observed in the individual 3D plots.}
	\label{fig:3}
	\vspace{-0.75cm}
\end{figure*}

The observed diversity of the ground states suggests various magnetization scenarios at zero temperature. In fact, the zero-temperature magnetization curves of the studied spin-$1/2$ Ising-Heisenberg orthogonal-dimer chain may exhibit the zero plateau as well as intermediate magnetization plateaus at one-quarter, one-half and three-quarters of the saturation magnetization according to the Oshikawa-Yamanaka-Affleck rule~\cite{Osh97, Aff98} as long as the period doubling of a magnetic unit cell is considered. In accordance with this rule, the plateau at zero magnetization is pertinent either to SAF or MAF ground state, the intermediate 1/4-plateau corresponds to the MFI ground state, the intermediate 1/2-plateau relates either to the SFI or FI ground state, while the last possible intermediate 3/4-plateau does not emerge in general. A comprehensive view of the situation is provided by three-dimensional (3D) plots of the total magnetization $m$ normalized with respect to its saturation value $m_{sat}$, which are depicted  in Fig.~\ref{fig:3} against the magnetic field $h/J_{I}$ and the interaction ratio $J_{I}^{\prime}/J_{I}$ by assuming either zero temperature $k_{\rm B}T/J_{I} = 0$ (left panels) or sufficiently small but finite temperature $k_{\rm B}T/J_{I} = 0.1$ (right panels). For the sake of a comparison, the interaction ratio $J_{H}/J_{I}$ is fixed to the same values as used for a construction of the ground-state phase diagrams depicted in Figs.~\ref{fig:2}b-d. Obviously, the zero-temperature magnetization curves plotted in left panels of Fig.~\ref{fig:3} reflect up to six different sequences of the field-driven phase transitions depending on a mutual interplay between the intra- and inter-dimer coupling constants $J_{H}$, $J_{I}$, $J_{I}^{\prime}$:
\begin{itemize}
	\item[(i)] MAF $\to$ SFI $\to$ FM\,,
	\item[(ii)] MAF $\to$ MFI $\to$ SFI $\to$ FM\,,
	\item[(iii)] MAF $\to$ MFI $\to$ FI $\to$ FM\,,
	\item[(iv)] SAF $\to$ MFI $\to$ FI $\to$ FM\,,
	\item[(v)] SAF $\to$ FI $\to$ FM\,,
	\item[(vi)] SAF $\to$ MFI $\to$ SFI $\to$ FM\,.
\end{itemize} 
In agreement with the aforementioned ground-state analysis, the sequences of the field-induced phase transitions MAF $\to$ SFI $\to$ FM, MAF $\to$ MFI $\to$ SFI $\to$ FM, SAF $\to$ MFI $\to$ FI $\to$ FM and SAF $\to$ FI $\to$ FM emerge in the zero-temperature magnetization curves for any value of the interaction ratio $J_{H}/J_{I}$, while the ones MAF $\to$ MFI $\to$ FI $\to$ FM and  SAF $\to$ MFI $\to$ SFI $\to$ FM can be identified  in the magnetization process only for $J_{H}/J_{I}<\sqrt{2}/2$ and $J_{H}/J_{I}>\sqrt{2}/2$, respectively (see Figs.~\ref{fig:3}a1 and~\ref{fig:3}c1). Of course, the actual magnetization plateaus and discontinuous magnetization jumps at the critical fields, which correspond to individual field-induced phase transitions, can be detected merely at zero temperature, because any finite temperature completely wipes a discontinuity in the magnetization and also diminishes the perfect plateaus from the respective isothermal magnetization curves (see right panels of Fig.~\ref{fig:3}). In general, the staircase character of the magnetization curves is gradually smoothing upon increasing of temperature until it completely vanishes due to sufficiently strong thermal fluctuations. 

\subsection{Bipartite quantum entanglement}
\label{subsec:3.3}

To gain a better insight into a degree of bipartite quantum entanglement emergent within the pure states of the horizontal Heisenberg spin dimers, the zero-temperature phase diagrams displayed in Fig.~\ref{fig:2} are supplemented with the corresponding density plots of the concurrence calculated according to Eq.~(\ref{eq:C}). As expected, the concurrence $C$ becomes non-zero in all quantum ground states SAF, MAF, SFI, and MFI, while it equals zero within two classical ground states FM and FI phases with a full alignment of the horizontal Heisenberg dimers towards the magnetic field [see the respective eigenvectors~(\ref{eq:FM}) and (\ref{eq:FI})]. The quantum phases SAF, SFI, MAF and MFI generally show a different strength of the bipartite quantum entanglement as evidenced by zero-temperature asymptotic values of the concurrence:
\begin{eqnarray}
C_{\rm SAF} \!\!\!&=&\!\!\! C_{\rm SFI} = 1\,,\qquad 
C_{\rm MAF} = \frac{J_{H}/J_{I}}{\sqrt{4 + (J_{H}/J_{I})^{2}}}\,,
\nonumber\\
\label{eq:C_SAF,MAF, MFI}
C_{\rm MFI} \!\!\!&=&\!\!\! \frac{J_{H}/J_{I}}{\sqrt{1 + (J_{H}/J_{I})^{2}}}\,. 
\end{eqnarray}
It can be understood from Eq.~(\ref{eq:C_SAF,MAF, MFI}) as well as the density plots shown in Fig.~\ref{fig:2} that the Heisenberg dimers residing on the horizontal bonds are fully entangled only within the SAF and SFI ground states. On the other hand, the strength of the bipartite quantum entanglement within the MAF and MFI ground states basically depends on a relative strength of the Heisenberg and Ising intra-dimer interactions. More specifically, the higher value the interaction ratio $J_{H}/J_{I}$ takes, the more strongly entangled the horizontal Heisenberg dimers are. It could be generally concluded that the Heisenberg dimers generally display a stronger quantum entanglement in the MFI phase than in the MAF phase when assuming the same value of the interaction ratio $J_{H}/J_{I}$. In contrast to the ground states SAF and SFI with a perfect quantum entanglement of the horizontal Heisenberg dimers, the perfect bipartite quantum entanglement cannot be reached neither in MAF nor in MFI phase for any finite value of the interaction ratio $J_{H}/J_{I}$, because the Heisenberg spin pairs reside in the outstanding singlet-like states instead of a perfect singlet-dimer state [cf. the eigenvectors~(\ref{eq:MFI}) and~(\ref{eq:MAF}) with the ones (\ref{eq:SFI}) and~(\ref{eq:SAF})]. 

\begin{figure*}[th!]
	\centering
	\vspace{0.0cm}
	\includegraphics[width=1.0\textwidth]{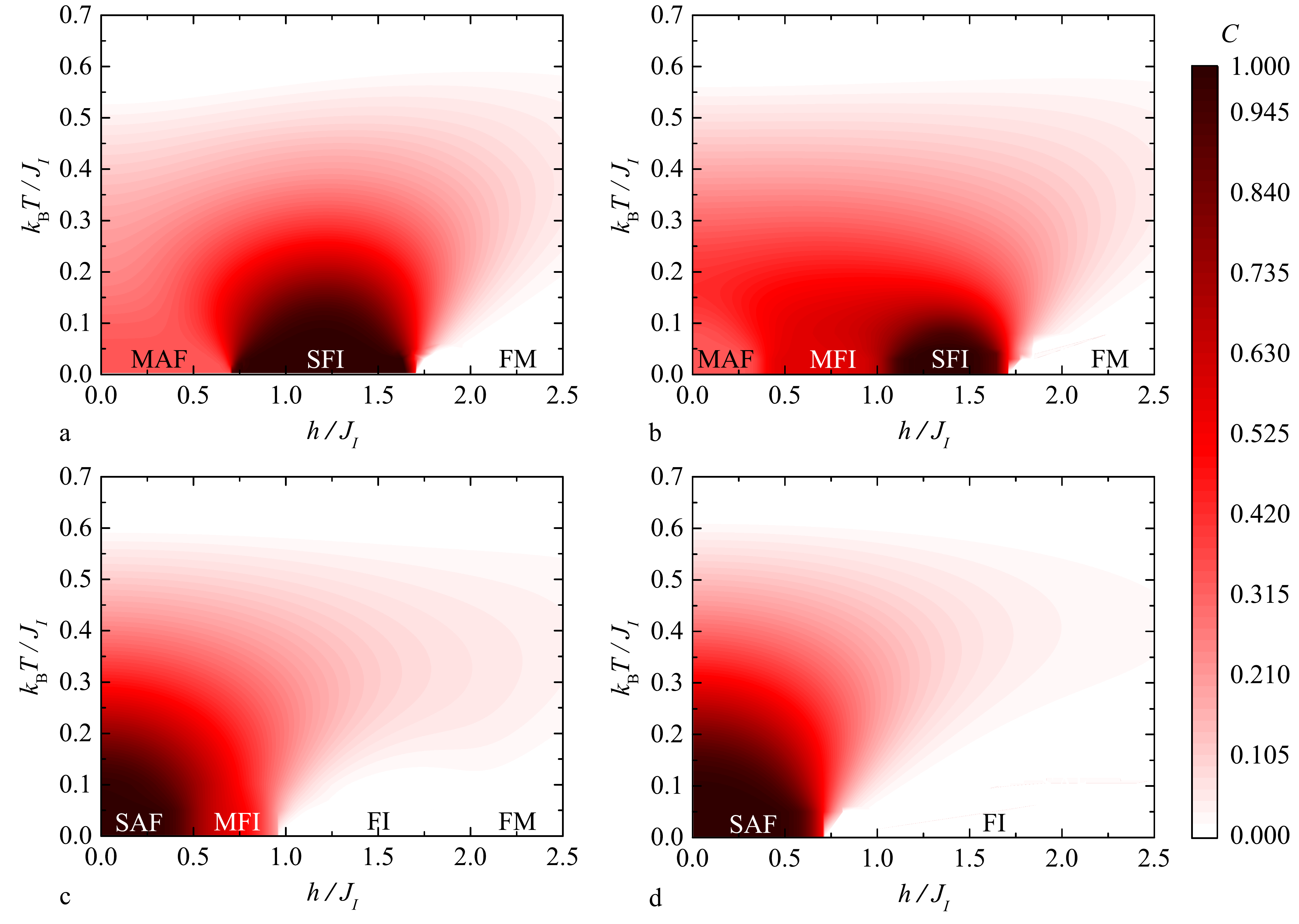}
	\vspace{-0.5cm}
	\caption{The density plots of the concurrence $C$ in the $h/J_{I} - k_{\rm B}T/J_{I}$ parameter plane for the fixed values of the interaction ratio $J_{H}/J_{I}=\sqrt{2}/2$ and $J_{I}^{\prime}/J_{I}=1/4$ (panel a), $J_{I}^{\prime}/J_{I}=1$ (panel b), $J_{I}^{\prime}/J_{I}=2$ (panel c), $J_{I}^{\prime}/J_{I}=3$ (panel d).}
	\label{fig:4}
	\vspace{0.0cm}
\end{figure*}

\subsection{Bipartite thermal entanglement}
\label{subsec:3.4}

\begin{figure*}[t!]
	\centering
	\vspace{0.0cm}
	\includegraphics[width=1.0\textwidth]{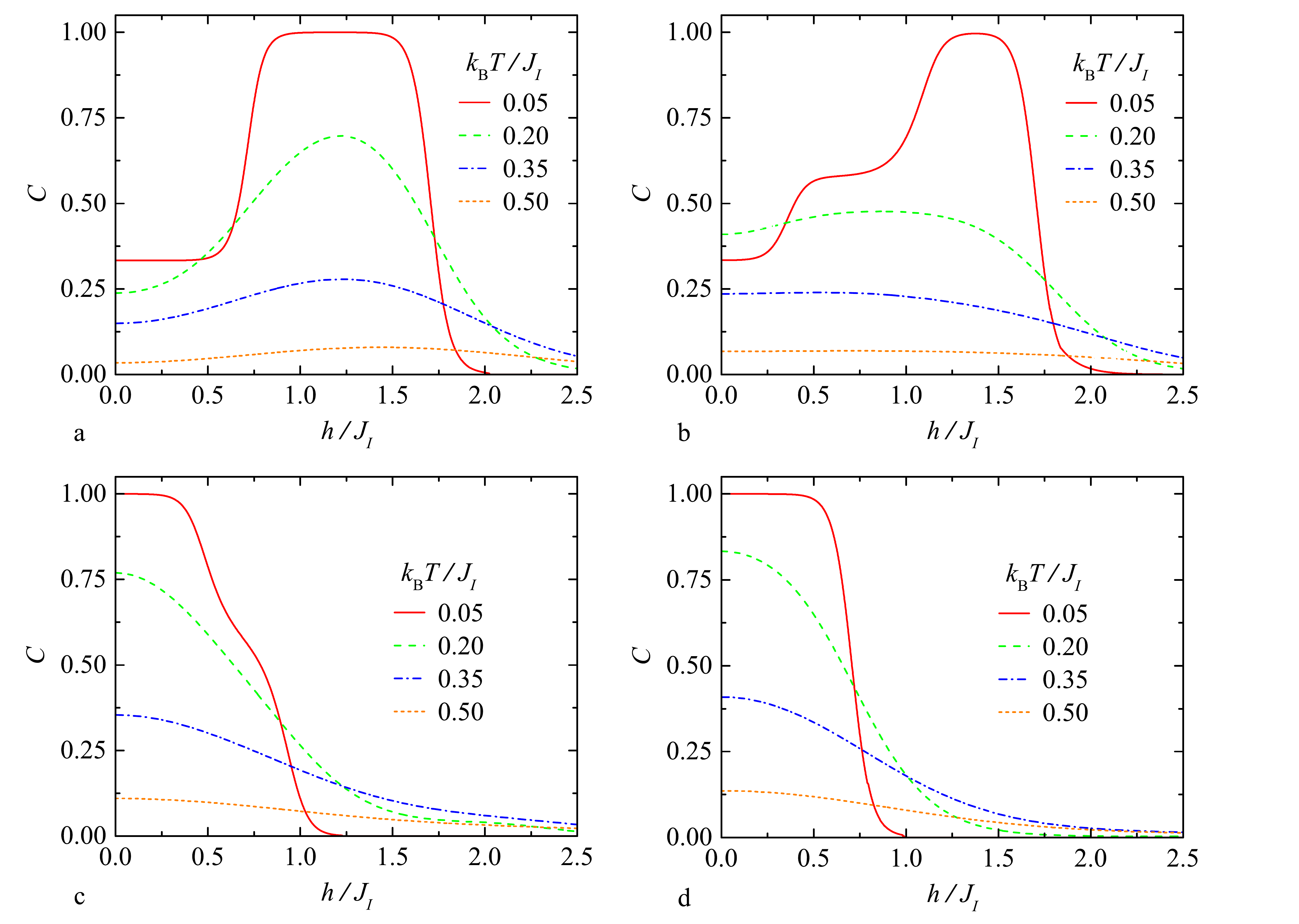}
	\vspace{-0.5cm}
	\caption{The magnetic-field dependencies of the concurrence $C$ for the fixed values of the interaction ratio $J_{H}/J_{I}=\sqrt{2}/2$ and $J_{I}^{\prime}/J_{I}=1/4$ (panel~a), $J_{I}^{\prime}/J_{I}=1$ (panel~b), $J_{I}^{\prime}/J_{I}=2$ (panel~c), $J_{I}^{\prime}/J_{I}=3$ (panel~d) by assuming four different values of temperature $k_{\rm B}T/J_{I}$.}
	\label{fig:5}
		\centering
	\vspace{0.75cm}
	\includegraphics[width=1.0\textwidth]{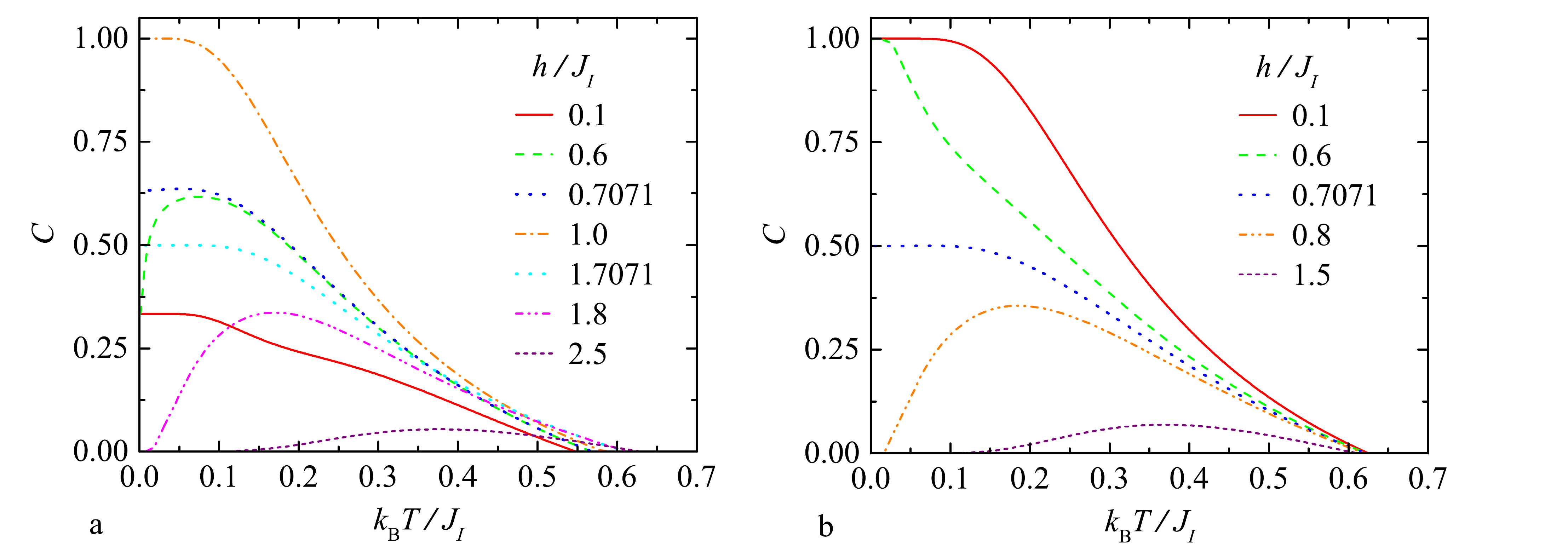}
	\vspace{-0.5cm}
	\caption{The temperature dependencies of the concurrence $C$ for the fixed values of the interaction ratio $J_{H}/J_{I}=\sqrt{2}/2$ and $J_{I}^{\prime}/J_{I}=0.25$ (panel a), $J_{I}^{\prime}/J_{I}=3$ (panel b) by assuming a few different values of the external magnetic field $h/J_{I}$.}
	\label{fig:6}
	\vspace{0.0cm}
\end{figure*}

Last but not least, let us turn our attention to a detailed examination of the bipartite thermal entanglement, which refers to a bipartite entanglement emergent within the mixed states of the horizontal Heisenberg dimers at nonzero temperatures. The overall picture of this issue can easily be created from density plots of the concurrence $C$ along with its magnetic-field and temperature dependencies, which are depicted in Figs.~\ref{fig:4}--\ref{fig:6} for the particular value of the interaction ratio  $J_{H}/J_{I}=\sqrt{2}/2$ and four different values of the interaction ratio $J_{I}^{\prime}/J_{I} = 1/4, 1, 2$ and $3$. As one could expect, the displayed data for the concurrence at low enough temperatures $k_{\rm B}T/J_{I} \lesssim 0.05$ faithfully resemble zero-temperature asymptotic values, which were discussed in above by the ground-state analysis. For the interaction ratios $J_{I}^{\prime}/J_{I} = 1/4$ and $1$ the concurrence $C$ first increases upon increasing of the magnetic field near the critical fields $h_{c1}/J_{I} \approx 0.7071$ (for $J_{I}^{\prime}/J_{I} = 1/4$) and $h_{c1}/J_{I} \approx 0.3966$,  $h_{c2}/J_{I} \approx 1.0176$ (for $J_{I}^{\prime}/J_{I} = 1$) due to a strengthening of the bipartite entanglement within the horizontal dimers near the field-induced phase transitions ${\rm MAF}\to{\rm SFI}$ and ${\rm MAF}\to{\rm MFI}$, ${\rm MFI}\to{\rm SFI}$, respectively. The second (third) field-induced phase transition ${\rm SFI}\to{\rm FM}$, which can be found for both particular values of the interaction ratio $J_{I}^{\prime}/J_{I}$ at the same saturation field $h_{c2(3)}/J_{I} \approx 1.7071$ is responsible for a sudden drop of the concurrence $C$ to zero, which confirms a breakdown of the bipartite entanglement (see Figs.~\ref{fig:4}a,b and~\ref{fig:5}a,b). On the other hand, the bipartite entanglement of the horizontal Heisenberg dimers generally weakens for $J_{I}^{\prime}/J_{I} = 2$ along the whole magnetization process. In fact, the first rapid decrease of the concurrence observable around the critical field $h_{c1}/J_{I} \approx 0.4824$ is attributable to the field-induced transition ${\rm SAF}\to{\rm MFI}$, while the second abrupt decline of the concurrence associated with a complete breakdown of the concurrence emerges near the critical field $h_{c2}/J_{I} \approx 0.9319$ of the field-driven phase transition ${\rm MFI}\to{\rm FI}$ (see Figs.~\ref{fig:4}c and~\ref{fig:5}c).
Finally, the bipartite thermal entanglement disappears upon strengthening of the magnetic field according to the most standard scheme for the highest value of the interaction ratio $J_{I}^{\prime}/J_{I} = 3$ illustrated in Figs.~\ref{fig:4}d and~\ref{fig:5}d. A breakdown of the concurrence already appears in a vicinity of the first critical field $h_{c1}/J_{I} \approx 0.7071$, which corresponds to the field-induced phase transition from the quantum ground state ${\rm SAF}$ to the classical one ${\rm FI}$ (see Fig.~\ref{fig:4}d and also Fig.~\ref{fig:5}d).

Of course, an increase in temperature causes a gradual \linebreak smoothing of abrupt changes of the concurrence observable at low enough temperatures in a proximity of the critical fields, because the bipartite entanglement between the Heisenberg spin pairs is in general suppressed by thermal fluctuations above all quantum ground states (see Fig.~\ref{fig:5}). However, it should be also mentioned that a small temperature rise may eventually invoke a gentle strengthening of the thermal entanglement. As a matter of fact, the concurrence $C$ may exhibit an outstanding temperature-induced rise on account of thermal excitations from a less entangled quantum ground state towards a more entangled excited state (see for instance the low-field parts of solid red and dashed green curves corresponding to the temperatures $k_{\rm B}T/J_{I} = 0.05$ and $0.2$ in Figs.~\ref{fig:5}a,b and the dashed green curve for the magnetic field $h/J_{I} = 0.6$ in Fig.~\ref{fig:6}a. Furthermore, the density plots of the concurrence along with the magnetic-field and temperature dependencies depicted in Figs.~\ref{fig:4}--\ref{fig:6} clearly evidence that the thermal entanglement of the horizontal Heisenberg dimers, although relatively weak, is also detectable at nonzero temperatures above the classical FI and FM ground states. This peculiar finding can be repeatedly explained in terms of a thermal activation of the entangled low-lying excited states related to some of the quantum phases SFI, MFI, MAF or SAF. 

\section{Conclusion}
\label{sec:4}

In the present work we have introduced and exactly solved a spin-1/2 Ising-Heisenberg orthogonal-dimer chain, which is composed of regularly alternating Ising and Heisenberg dimers placed in an external magnetic field. After tracing out spin degrees of freedom of the Heisenberg dimers, the considered quantum spin chain has been rigorously treated by making use of the classical transfer-matrix approach. It is shown that the ground-state phase diagram involves in total six different ground states. In addition to the classical ferromagnetic phase emergent above the saturation field one also encounters two ground states with zero total magnetization referred to as the singlet and modulated antiferromagnetic phases, two ground states with the total magnetization equal to a half of the saturation value referred to as the frustrated ferrimagnetic phase and the singlet ferrimagnetic phase and, finally, one peculiar ground state with the total magnetization equal to a quarter of the saturation value referred to as the modulated ferrimagnetic phase. It is also evidenced that the diversity of the ground states gives rise to six different magnetization scenarios depending on a mutual interplay of three considered coupling constants.

A particular attention has been paid to quantification of the bipartite entanglement within the pure and mixed states of the horizontal Heisenberg dimers at zero and nonzero temperatures with the help of concurrence. Surprisingly, it turns out that the bipartite entanglement may be reinforced by increasing of temperature or even upon strengthening of the magnetic field, which is in contrast with general expectations. In addition, the bipartite thermal entanglement has been identified also above two classical phases: the ferromagnetic phase and the frustrated ferrimagnetic phase. This unexpected finding can be ascribed to thermal excitations driving the investigated quantum spin chain from a pure classical ground state to a mixed state incorporating low-lying excited state(s) closely connected to other quantum ground states. 

Although the magnetic structure of the investigated spin-1/2 Ising-Heisenberg  orthogonal-dimer chain was inspired by a heterobimetallic backbone of the coordination polymer [Dy$_2$Cu$_2$]$_n$, the substantial difference between the Land\'e g-factors of Dy$^{3+}$ and Cu$^{2+}$ magnetic ions ($g_{\rm Dy} \approx 20$ vs. $g_{\rm Cu} \approx 2.2$) precludes a straightforward comparison of the obtained theoretical results with the available experimental data \cite{ueki2007,okazawa2008}. The investigation of the difference of the respective Land\'e g-factors along with anisotropy of the couplings constants is accordingly left as future task for our forthcoming study.  

\section*{Acknowledgment}
This work was financially supported by the grant of The Ministry of Education, Science, Research and Sport of the Slovak Republic under the contract No. VEGA 1/0105/20 and by the grant of the Slovak Research and Development Agency under the contract No. APVV-16-0186.


\end{document}